\documentclass[letterpaper,conference]{IEEEtran}
\usepackage[utf8]{inputenc}
\usepackage{cite}
\usepackage{amsmath,amssymb,esint}
\usepackage{graphicx,xcolor}
\usepackage{multirow,array,dblfloatfix}
\usepackage[super]{nth}
\usepackage{psfrag}
\usepackage{pbox,ragged2e}
\usepackage{balance}

\usepackage{graphicx}
\usepackage[caption=false,font=footnotesize]{subfig}

\newcommand{\f}[2]{\frac{#1}{#2}}

\newcommand{\mtx}[2]{\left[\begin{array}{#1}#2\end{array}\right]}
\newcommand{\Tr}{^\mathrm{T}}

\newcommand{\eye}{\mathbf{I}}

\newcommand{\ttx}{\mathrm{T}}
\newcommand{\trx}{\mathrm{R}}
\renewcommand{\t}{_\ttx}
\renewcommand{\r}{_\trx}
\newcommand{\tr}{_{\ttx\trx}}

\newcommand{\Z}{\mathbf{Z}}

\DeclareMathOperator{\re}{Re}
\DeclareMathOperator{\im}{Im}

\begin{document}%
\title{Magneto-inductive Passive Relaying\\ in Arbitrarily Arranged Networks}%
%\author{%
%\IEEEauthorblockN{Gregor Dumphart, Eric Slottke, and Armin Wittneben} 
%\IEEEauthorblockA{Communication Technology Laboratory, ETH Zurich, 8092 Zurich, Switzerland\\
%Email: \{dumphart, wittneben\}@nari.ee.ethz.ch}}
\author{%
\IEEEauthorblockN{Gregor Dumphart, Eric Slottke, and Armin Wittneben}%
\IEEEauthorblockA{Communication Technology Laboratory, ETH Zurich, 8092 Zurich, Switzerland\\
Email: \{dumphart,  slottke, wittneben\}@nari.ee.ethz.ch}}%
\maketitle
\begin{abstract}
We consider a wireless sensor network that uses inductive near-field coupling for wireless powering or communication, or for both. The severely limited range of an inductively coupled source-destination pair can be improved using resonant relay devices, which are purely passive in nature. Utilization of such magneto-inductive relays has only been studied for regular network topologies, allowing simplified assumptions on the mutual antenna couplings. In this work we present an analysis of magneto-inductive passive relaying in arbitrarily arranged networks. We find that the resulting channel has characteristics similar to multipath fading: the channel power gain is governed by a non-coherent sum of phasors, resulting in increased frequency selectivity. We propose and study two strategies to increase the channel power gain of random relay networks: \mbox{i) deactivation} of individual relays by open-circuit switching and ii) frequency tuning. The presented results show that both methods improve the utilization of available passive relays, leading to reliable and significant performance gains.
\end{abstract}
\section{Introduction}
\label{sec:intro}
Inductive coupling is an established wireless physical layer technology based on the exchange of energy in the magnetic near-field, well known from its application in RFID and NFC standards \cite{Finkenzeller2010}. Such magneto-inductive (MI) systems are characterized by low complexity requirements, the ability to efficiently supply power wirelessly, and robustness to harsh propagation media, which led to the proposal of MI communications and powering for areas such as sensor networks \cite{loh2008inductively}, biomedical implants \cite{Campi2016}, or the internet of things \cite{mattern2010internet}.

\begin{figure}[!t]
	\centering
  \includegraphics[width=.72\linewidth,trim=9 0 0 0,clip=true]{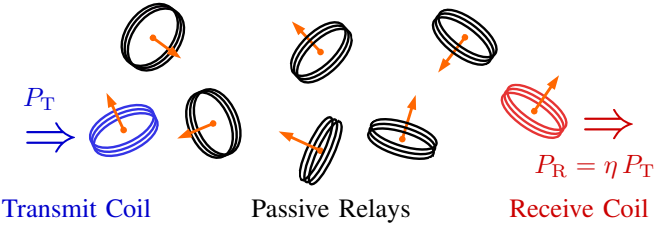} % Above trim: 115 is kinda sharp
  \put(-206,13){\color[rgb]{.0,.0,.8}\huge$\Rightarrow$}
	%\put(-214.2,13){\color[rgb]{.0,.0,.8}$P\t$}
	\put(-206,  30){\color[rgb]{.0,.0,.8}$P\t$}
	%\put( - 8, 43){\color[rgb]{.8,.0,.0}\Large$\Longrightarrow$}
	\put(    5, 18){\color[rgb]{.8,.0,.0}\huge$\Rightarrow$}
	\put(-12.5,4.5){\color[rgb]{.8,.0,.0}$P\r = \eta \, P\t$}
  \put(-214  ,-11.9){\color[rgb]{.0,.0,.8}Transmit Coil}
	%\put(-119.6,-11.9){\color[rgb]{.39,.39,.39}Passive Relays}
	\put(-119.6,-11.9){Passive Relays}
	\put( -22  ,-11.9){\color[rgb]{.8,.0,.0}Receive Coil}
	\caption{Passive MI relaying network of arbitrary arrangement. The link from transmitter to receiver is assisted by the presence of resonant coils (i.e. passive relays) which may increase the channel power gain $\eta$.}
	\label{fig:IntroFig}
	\vspace{-6mm}
\end{figure}

However, an important drawback of MI systems is their small usable range, which is due to the $d^{-3}$ decay of magnetic near-field magnitude with distance $d$. A promising approach to overcome this range limitation is the concept of MI passive relaying, where the link is improved by one or multiple relay devices present in the vicinity of transmitter (Tx) and receiver (Rx), as shown in Fig.~\ref{fig:IntroFig}. The relays, which are implemented as purely passive resonant circuits, are hereby excited by the Tx field and subsequently generate a secondary magnetic field which couples to other relays and the Rx. The concept is formalized in Sec.~\ref{sec:model}. 

The notion of MI relaying has been proposed by Shamonia et al. as a novel type of waveguide \cite{shamonina2002magneto}, in which the relay elements have been regularly arranged between Tx and Rx in a coaxial fashion. We, however, envision a cooperative scheme in dense inductively coupled sensor networks, by which idle nodes may act as relays to improve the channel between their currently operating counterparts. As sensor networks are typically mobile or deployed in an ad-hoc fashion, the node locations and orientations can be considered arbitrary. However, to the best of the authors' knowledge, MI relaying for wireless powering or communication has only been considered for regular arrangements, allowing simplifying assumptions on the node couplings \cite{shamonina2002magneto,sun2012capacity,LeeTPE2012,KisseleffTC2014,chan2011two}. 
%\textcolor[rgb]{0.92,0.5,0.4}{(Hier wirklich alle verwendeten Referenzen anführen, die das erfuellen)}
For example, the channel capacity is derived in \cite{sun2012capacity} for 1-D networks of equidistant, coplanar relays, considering only couplings between neighboring coils. The work contains an analysis of failure or misplacement of a single relay.
The effect of coupling between non-adjacent relays is studied in \cite{LeeTPE2012} for wireless power transfer.
In \cite{KisseleffTC2014} the throughput of underground MI relaying networks with spanning tree topology is optimized, which includes the adjustment of coil orientations for interference zero-forcing. 
The authors of \cite{chan2011two} analyze MI communication over a 2-D grid of relays.

In this work, we present the first analysis of passive MI relaying networks with arbitrary arrangements. To this end, we develop a generalized system model for MI links in the presence of passive relays and discuss the impact of irregular arrangement on the channel power gain. We show by simulation that randomly deployed relays can both significantly increase or decrease the power gain, depending on the density and individual geometry realization of the network. This behavior similar to multipath fading is primarily caused by a non-coherent superposition of individual relay contributions to the effective link transimpedance. This limits the practical merits of random networks to the compensation of misalignment losses, whereas significant further gains are improbable. For better utilization  of the relaying channel we propose an optimization scheme based on the deactivation of individual relays by load switching, which leads to considerable and reliable improvements and can therefore be considered a key enabler for wireless power transfer and communication in ad-hoc inductively coupled sensor networks. We identify frequency tuning as a reasonable low-complexity alternative.

\section{Generalized Model for MI Relaying}
\label{sec:model}
\begin{figure}[!ht]
\centering
\psfrag{M}{\hspace{-1.7mm}$M\tr$}
\psfrag{Rt}{$R\t$}
\psfrag{Rr}{$R\r$}
\psfrag{Lt}{$L\t$}
\psfrag{Lr}{$L\r$}
\psfrag{ZS}{\raisebox{-.5mm}{\hspace{1mm}\color[rgb]{0,0,.8}$Z_\text{in}^*$}}
\psfrag{ZL}{\hspace{-.5mm}\raisebox{-.5mm}{\color[rgb]{.8,0,0}$Z_\text{out}^*$}}
\psfrag{ZN}{\hspace{-0mm}\raisebox{.7mm}{$\Z\tr$}}
%\psfrag{MS}{\color[rgb]{.0,.0,.8}Matched Source}
\psfrag{MS}{\hspace{4mm}\color[rgb]{0,0,.8}\pbox{2cm}{\RaggedRight Matched\\Tx Source}}
%\psfrag{IC}{\hspace{-2.1mm}\raisebox{1.8mm}{\color[rgb]{.38,.38,.38}Inductive Coupling}}
\psfrag{IC}{\hspace{-2.1mm}\raisebox{1.8mm}{Inductive Coupling}}
\psfrag{ML}{\hspace{7mm}\color[rgb]{.8,0,0}\pbox{2cm}{\RaggedLeft Matched\\Rx Load}}
\psfrag{At}{\hspace{-1.3mm}\color[rgb]{0,0,.8}\large$\Longrightarrow$}
%\psfrag{At}{\hspace{-1.3mm}\color[rgb]{0,0,.8}\huge$\Rightarrow$}
\psfrag{Pt}{\hspace{-1mm}\color[rgb]{0,0,.8}$P\t$}
\psfrag{Ar}{\hspace{-.3mm}\color[rgb]{.8,0,0}\large$\Longrightarrow$}
\psfrag{Pr}{\color[rgb]{.8,0,0}$P\r$}
\includegraphics[width=\linewidth,trim=13 0 7 2,clip=true]{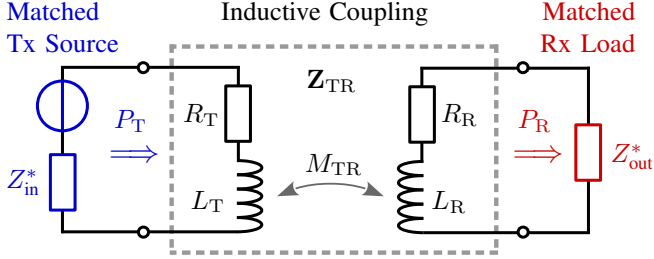} % best fit to text: trim=18 0 12 2
\caption{Circuit model of a conventional MI link. The coils are coupled through $M\tr$, which is modeled by a two-port network of impedance matrix $\Z\tr$.}
\label{fig:direct_channel} 
\end{figure}

In this section, we introduce a general model of the MI relaying channel and give a formula for the power gain that can be achieved over this channel. The model follows by extension of an established circuit-theoretic  description of an inductively coupled link between a Tx-Rx pair \cite{Xue2013}. We start with a review of this simpler case, which is depicted in Fig.~\ref{fig:direct_channel}. 

The Tx and Rx nodes are equipped with near-field loop antennas (i.e. coils) which are characterized by self-inductances $L\t, L\r$ in series with resistances $R\t, R\r$ accounting for ohmic and radiative losses. The antenna near-field coupling is described by the mutual inductance $M\tr = k\tr \sqrt{L\t L\r}$ with coupling coefficient $k\tr \in \left[-1, +1\right]$. $M\tr$ depends on the separation and relative arrangement of the coils as well as their individual geometries \cite{DumphartPIMRC2016}. The impedance matrix of the two-port network formed by the coupled coils is given as
\begin{align}
\Z\tr
&= \mtx{cc}{R\t+j\omega L\t & j\omega M\tr \\ j\omega M\tr & R\r+j\omega L\r}.
\label{eq:ZDirect}
\end{align}

Maximum power transfer is achieved by choosing the Tx source and the Rx load such that conjugate matching with the respective encountered antenna impedance is established \cite{Pozar2004}. In this simultaneous matching problem \cite{Rahola2008}, each termination affects the opposite antenna impedance through $M\tr$.\footnotemark
\footnotetext{For fixed $\omega$ and $M\tr$, the required conjugate impedances can be established by transforming the actual termination impedances (e.g., $50\,\Omega$) with lossless matching two-ports of reactive lumped elements \cite{Xue2013,Pozar2004}, leading to resonance at $\omega$ with the typical capacitor. See the appendix for port impedance formulas. For non-constant $M\tr$ or over a frequency band, the matching could be established with adaptive or broadband matching networks, respectively. We assume simultaneous conjugate matching to study the potential of MI relaying channels by means of $\eta$, which can be regarded as an upper bound that can be achieved with appropriate matching networks.}
The Tx source delivers active power $P\t$ to the Tx coil while the active power into the Rx load is $P\r = \eta P\t$. We refer to $\eta$ as channel power gain and use it as a measure of link quality.\footnotemark
\footnotetext{The channel power gain $\eta$ is equivalent to the squared channel gain $|h|^2$ in a communications context and maximum power transfer efficiency in a wireless power transfer context.}
Its value for the discussed conventional MI link is known in the literature (refer to \cite{Xue2013} for example) to be 
\begin{align}
\eta = \f{\chi\tr^2}{\big(1+\sqrt{1+\chi\tr^2} \, \big)^2}
\label{eq:PowerGainDirect} \ ,
\end{align}
where we have introduced the shorthand definition
\begin{align}
\chi\tr = \f{\omega M\tr}{\sqrt{R\t R\r}} = k\tr \sqrt{Q\t Q\r} \ .
\label{eq:betaDirect}
\end{align}
The power gain depends solely on the coupling coefficient $k\tr$ and the respective $Q$-factors of the antennas, $Q\t = \omega L\t / R\t$ and $Q\r = \omega L\r / R\r$. It can be noted that for $\chi\tr \gg 1$ we observe $\eta\approx 1$, i.e. the MI channel becomes effectively lossless. In contrast, we encounter $\eta \approx \f{1}{4} \chi\tr^2$ when $\chi\tr$ is small because of weak coupling.

\begin{figure}[!ht]
\centering
\psfrag{M}{\hspace{-1.7mm}$M\tr$}
\psfrag{Rt}{$R\t$}
\psfrag{Rr}{$R\r$}
\psfrag{Lt}{$L\t$}
\psfrag{Lr}{$L\r$}
\psfrag{ZS}{\raisebox{-.5mm}{\hspace{1mm}\color[rgb]{0,0,.8}$Z_\text{in}^*$}}
\psfrag{ZL}{\hspace{-.5mm}\raisebox{-.5mm}{\color[rgb]{.8,0,0}$Z_\text{out}^*$}}
\psfrag{Z}{\hspace{-0mm}\raisebox{.7mm}{$\Z$}}
%\psfrag{MS}{\color[rgb]{.0,.0,.8}Matched Source}
\psfrag{MS}{\hspace{4mm}\color[rgb]{0,0,.8}\pbox{2cm}{\RaggedRight Matched\\Tx Source}}
%\psfrag{IC}{\hspace{-5.8mm}\raisebox{1.8mm}{\color[rgb]{.38,.38,.38}Passive Relaying Channel}}
\psfrag{IC}{\hspace{-5.8mm}\raisebox{1.8mm}{Passive Relaying Channel}}
\psfrag{ML}{\hspace{7mm}\color[rgb]{.8,0,0}\pbox{2cm}{\RaggedLeft Matched\\Rx Load}}
\psfrag{At}{\hspace{-1.3mm}\color[rgb]{0,0,.8}\large$\Longrightarrow$}
\psfrag{Pt}{\hspace{-1mm}\color[rgb]{0,0,.8}$P\t$}
\psfrag{Ar}{\hspace{-.3mm}\color[rgb]{.8,0,0}\large$\Longrightarrow$}
\psfrag{Pr}{\color[rgb]{.8,0,0}$P\r$}
\psfrag{MN1}{\hspace{0.4mm}$M_{N,1}$}
\psfrag{MT1}{\hspace{0.8mm}$M_{\ttx,1}$}
\psfrag{MTN}{\hspace{0.3mm}$M_{\ttx,N}$}
\psfrag{MR1}{$M_{\trx,1}$}
\psfrag{MRN}{$M_{\trx,N}$}
\psfrag{R1}{$R_1$}
\psfrag{RN}{$R_N$}
\psfrag{L1}{$L_1$}
\psfrag{LN}{$L_N$}
\psfrag{dots}{\hspace{2.4mm}\raisebox{.5mm}{$\cdots$}}
%\psfrag{Z1}{\raisebox{.5mm}{$Z_1$}}
%\psfrag{ZN}{\raisebox{.5mm}{$Z_N$}}
\psfrag{Z1}{$\tilde{Z}_1$}
\psfrag{ZN}{$\tilde{Z}_N$}
\includegraphics[width=\linewidth,trim=13 0 7 2,clip=true]{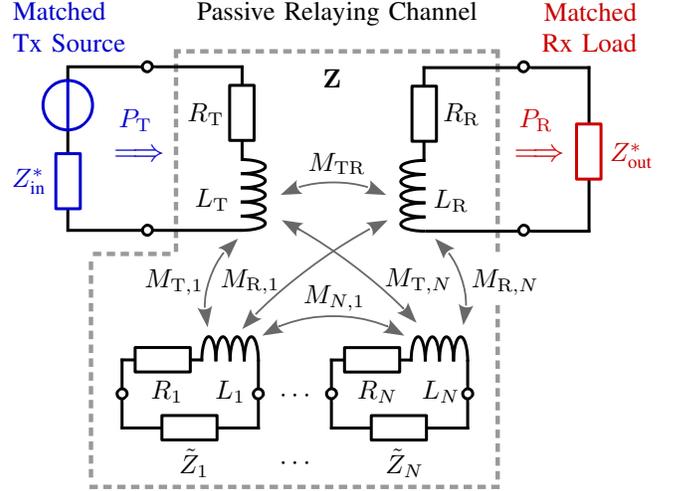} % best fit to text: trim=18 0 12 2
\caption{Circuit model of a MI link with $N$ passive relays.}
\label{fig:relay_channel}
\end{figure}

We will now extend the discussed MI link model by the introduction of $N$ passive relaying devices as shown in Fig.~\ref{fig:relay_channel}. 

The $n$-th relay hereby is a coil antenna with inductance $L_n$ and loss resistance $R_n$. It is terminated by a load impedance $\tilde{Z}_n \in \mathbb{C}$ which can be frequency-dependent. As the relays are passive, we require $\re \tilde{Z}_n \geq 0$. 
Following the original proposal of MI relaying, we design the relays resonantly at $\omega_0$ by using capacitors as relay loads, so $\tilde{Z}_n = 1/(j\omega C_n)$ with $C_n = 1/(\omega_0^2 L_n)$. Thereby, $\omega_0$ is the chosen operating frequency (i.e. the carrier frequency for communication).

We incorporate the action of the $N$ relays into the Tx-Rx two-port as follows: we consider the $(2+N)$-port network between all coils and terminate the $N$ relay ports by their individual loads $\tilde{Z}_n$, leaving two ports \cite{hassan2015}. 
We obtain a description of the passive MI relaying channel in terms of the $2 \times 2$ impedance matrix %of the effective Tx-Rx two-port network
\begin{align}
\Z = 
\Z\tr + \omega^2 \, \textbf{M} \, \Z_\mathrm{Rs}^{-1} \textbf{M}\Tr
\label{eq:EquivTwoport}
\end{align}
where we have introduced the matrices
\begin{align}
\Z_\mathrm{Rs} =& \mtx{l}{
\! R_1 \! + \! j\omega L_1 \! + \! \tilde{Z}_1 \ \ \ \ \ \ j\omega M_{2,1} \ \ \ \cdots \ \ \ j\omega M_{N,1} \\
\ \ j\omega M_{2,1} \ \ \ R_2 \! + \! j\omega L_2 \! + \! \tilde{Z}_2 \ \ \ddots \ \ \ \ \ \ \ \ \vdots  \\
\ \ \ \ \ \vdots \ \ \ \ \ \ \ \ddots \ \ \ \ \ \ \ \ \ \ \ \ \ \ \ddots \ \ \ \ \ j\omega M_{N,N-1} \\
\ \ j\omega M_{N,1} \ \cdots \ j\omega M_{N,N-1} \ \ R_N \! + \! j\omega L_N \! + \! \tilde{Z}_N \!\!
}, \nonumber \\
\textbf{M} =& \hspace{.3mm}\mtx{ccc}{M_{\ttx,1} & \dots & M_{\ttx,N} \\ M_{\trx,1} & \dots & M_{\trx,N}} .
\label{eq:ImpMatrices}
\end{align}
Note that the mutual inductance terms between each coil pair in the network again depend on their relative arrangement, which may be arbitrary. Again, we assume conjugate matching at both ends. In contrast to $\Z\tr$, the matrix $\Z$ will in general exhibit a transimpedance of non-zero real part $\re Z_{2,1}$. The channel power gain for this general case is given by
\begin{align}
\eta 
&= \f{\rho^2 + \chi^2}{\Big(\,\sqrt{1-\rho^2} + \sqrt{1+\chi^2}\,\Big)^2} 
\label{eq:PowerGain}
\end{align}
with $\rho \in [-1,+1]$ and $\chi \in \mathbb{R}$ defined as
\begin{align}
\rho &= \f{\re Z_{2,1}}{\sqrt{\re Z_{1,1}} \sqrt{\re Z_{2,2}}} \ , \! &
\chi &= \f{\im Z_{2,1}}{\sqrt{\re Z_{1,1}} \sqrt{\re Z_{2,2}}} \ .
\label{eq:alphabeta}
\end{align}
This generalization is derived and discussed in the appendix. Note that $\eta$ without relays \eqref{eq:PowerGainDirect} is included as special case, as choosing $\Z = \Z\tr$ results in $\rho = 0$ and $\chi = \chi\tr$.

\section{Analytic Study of MI Relaying}
\label{sec:study}
By verifying $\partial\eta/\partial(\rho^2) \geq 0$ and $\partial\eta/\partial(\chi^2) \geq 0$, we find that $\eta$ increases with $\rho^2$ and $\chi^2$. The maximum $\eta = 1$ is attained for $\rho^2 = 1$ or $\chi^2 \rightarrow \infty$. In other words, $\eta$ increases with 
\begin{align}
\rho^2 + \chi^2 = \f{|Z_{2,1}|^2}{\re Z_{1,1} \re Z_{2,2}} \ .
\label{eq:KeyToEta}
\end{align}
Thus, we will study the individual right-hand side terms of \eqref{eq:KeyToEta} in the following to provide an understanding of the merits and problems of passive relaying. First, we use \eqref{eq:EquivTwoport} to rewrite the crucial transimpedance term
\begin{align}
Z_{2,1} = j\omega M\tr + \omega^2 \sum_{n=1}^N \sum_{m=1}^N M_{\ttx,n} (\Z_\mathrm{Rs}^{-1})_{n,m} M_{\trx,m} \ .
\label{eq:SumOfPhasors}
\end{align}
The behavior of $Z_{2,1}$ is made intricate by $\Z_\mathrm{Rs}^{-1} \in \mathbb{C}^{N \times N}$ which accounts for the near-field interaction between all relay coils.
The elements $(\Z_\mathrm{Rs}^{-1})_{n,m} \in \mathbb{C}$ have general phase for an arbitrarily arranged network, making $Z_{2,1}$ a non-coherent sum of $1+N^2$ complex phasors. Hence, $|Z_{2,1}|$ may be large or close to zero (even for large $|M\tr|$), depending on how constructive or destructive the phasor sum turns out by virtue of the specific network arrangement. $Z_{2,1}$ is also highly frequency-dependent, which is especially due to $\Z_\mathrm{Rs}^{-1}$. The black graph in Fig.~\ref{fig:Spectrum} shows the frequency response of an exemplary relay channel.

\hyphenation{trans-impedance}
The non-coherent sum formulation \eqref{eq:SumOfPhasors} of transimpedance allows a more intuitive illustration of the passive relaying idea: the supporting relays provide alternative paths from Tx to Rx which may lead to a strong link even when $M\tr$ is weak, specifically when \mbox{$|Z_{2,1}| \gg |\omega M\tr|$}. Thus, the alternative paths are particularly helpful when $M\tr$ is small. The latter can be the result of misalignment (i.e. orientation-dependent attenuation typical in ad-hoc or mobile applications \cite{DumphartPIMRC2016}), in which case passive relays provide a means for increasing link reliability between randomly aligned nodes. 
On the other hand, when a well-aligned Tx-Rx pair shows weak $M\tr$ because of large separation, significant gains can be obtained with contiguous paths of strongly coupled relays from Tx to Rx. 
This is specifically realized by the regular arrangements studied in literature \cite{shamonina2002magneto,LeeTPE2012,sun2012capacity,KisseleffTC2014,chan2011two},
%\textcolor[rgb]{0.92,0.5,0.4}{(Hier wirklich alle verwendeten Referenzen anführen, die das erfuellen)}
but can also be achieved by general geometries.

The impact of $\re Z_{1,1}$ and $\re Z_{2,2}$ on $\eta$ can be explained by emphasizing their roles as open-circuit resistances of $\Z$: $\re Z_{1,1}$ is the encountered resistance of the Tx coil coupled with all relays but in the absence of the Rx device, and $\re Z_{2,2}$ vice versa. We observe $\re Z_{1,1} > R\t$ or $\re Z_{2,2} > R\r$ when a node is appreciably coupled to relays. Thus, increases in $\re Z_{1,1}$ or $\re Z_{2,2}$ indicate power dissipation in the lossy relay circuits (driven by induced currents), which decreases $\eta$.

%Example with two relays at $\omega = \omega_0$ schaut jetzt mords scheisse aus. Vielleicht sollten wir es ganz weglassen, weil es fälschlicherweise suggeriert a) dass höchstens 3-hop-Pfade betrachtet würden und auch b) dass die Summanden entweder rein imaginär oder rein reell wären. 
%\begin{multline}
%Z_{2,1} = j\omega M\tr + \f{\omega_0^2}{R_1 R_2 + \omega_0^2 M_{1,2}^2} \Big(
%R_2 M_{\ttx,1} M_{1,\trx} + \\ 
%R_1 M_{\ttx,2} M_{2,\trx} - j\omega_0 M_{\ttx,1} M_{1,2} M_{2,\trx} - j\omega_0 M_{\ttx,2} M_{2,1} M_{1,\trx} \Big)
%\end{multline}

In summary, we state the following conditions on passive relaying networks to improve $\eta$ of the Tx-Rx link: \mbox{i) $|Z_{2,1}|$} increases, i.e. the spatial relay density is adequate and does not lead to destructive phasor addition, and ii) power dissipation by the relay circuits does not outweigh the increase in $|Z_{2,1}|$.

While these insights explain important aspects of passive MI relaying, the behavior of arbitrarily arranged networks is still obscured by the analytical intractability of $\Z_\mathrm{Rs}^{-1}$ and the geometric dependencies of the mutual inductances. Thus, the next section presents a study based on simulation results.

%$\partial\eta/\partial(\rho^2) > 0$ and $\partial\eta/\partial(\chi^2) > 0$.
%Like in the direct link case, transimpedance is crucial for power transfer efficiency: $\eta$ increases with $|Z_{2,1}|$ and approaches $1$ for large $|Z_{2,1}|$. 

%We see that $\eta|_{\rho^2 = 0} \leq \eta \leq \eta|_{\rho^2 = 1} = 1$ for any $\chi$. Therefore, when $\chi^2$ becomes large, we see that $\eta$ approaches $1$ because $\eta|_{\rho = 0}$ already approaches $1$. Thus, $\eta$ increases with $Z_{2,1}$ bla bla

\section{Performance Evaluation for Arbitrary Arrangements}
\label{sec:eval}
We study the behavior of passive MI relaying quantitatively by considering random passive relaying networks and observing the resulting $\eta$ in comparison to the case without relays. 
The following parameters are used for all simulations. 
We assume that all involved nodes are equipped with identical flat circular coil antennas of $12\,\mathrm{mm}$ radius, twelve turns, self-inductance $L = 3.7\,\mu\mathrm{H}$, and serial resistance $R = 1\,\Omega$. 
All mutual inductances are calculated by numerical integration of the magnetoquasistatic Neumann formula for thin-wire loops \cite{neumann1846allgemeine}. 
% \cite{Griffiths1999} for submission.
% \cite{neumann1846allgemeine} for camera-ready.
The relays, terminated by identical load capacitors, are resonant at the chosen operating frequency $f_0 = 13.56\,\mathrm{MHz}$.\footnote{This is the ISM band used by standards ISO/IEC 14443 and 15693 \cite{Finkenzeller2010}.}

We assume a constant Tx-Rx distance of $50\,\mathrm{cm}$. Because of the role of $M\tr$ for relaying performance gains (as discussed in Sec.~\ref{sec:study}) we consider two cases of relative Tx-Rx coil arrangement: 
i) coaxial coils, constituting an ideally aligned pair, and ii) a misaligned pair, whose mutual inductance is attenuated to $10^{-23.7 / 20}$ times its coaxial value. This loss corresponds to the $10$th percentile value of $|M\tr|$ resulting from uniformly random coil orientations \cite{DumphartPIMRC2016}.
The relays are distributed randomly in three-dimensional space\footnotemark\ according to a uniform distribution, with an average volumetric relay density specified in $\mathrm{relays}/\mathrm{dm}^3$. All relay coil orientations are random with all possible directions being equiprobable. 
\footnotetext{We confine the relay network to a prolate spheroid whose foci are the Tx and Rx positions, respectively, and whose minor semi-diameter is equal to the Tx-Rx separation. This way, we limit the number of relays and computational cost, which for large $N$ is dominated by the inversion $\Z_\text{Rs}^{-1}$, while maintaining a homogeneous relay distribution between and around Tx and Rx.}

\begin{figure}[!ht]
	\centering
  \psfrag{CGdB}{\hspace{-13.5mm}\raisebox{-1.3mm}{\footnotesize{Channel Power Gain $\eta$ [$\mathrm{dB}$]}}}
  \psfrag{EmpCDF}{\hspace{-4.7mm}\raisebox{1.2mm}{\footnotesize{Empirical CDF}}}
	\includegraphics[width=\linewidth,trim=22 15 47 19,clip=true]{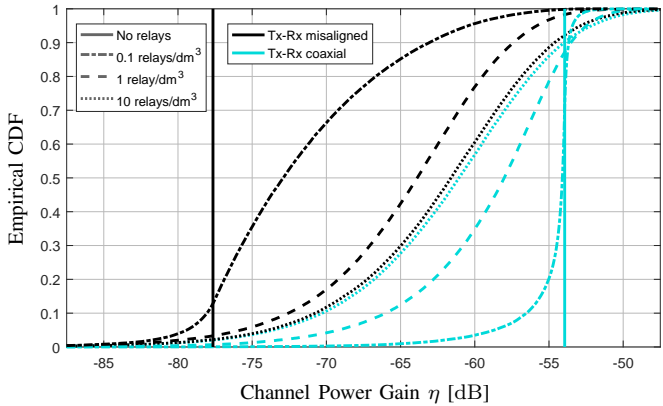}
  \vspace{-1.5mm}
  \caption{Statistics of the channel power gain between a Tx-Rx pair of $50\,\mathrm{cm}$ separation in a randomly arranged passive relaying network, evaluated for different volumetric relay densities. The vertical lines show $\eta$ without relays for two cases of relative Tx-Rx arrangement: i) an ideally arranged coaxial pair and ii) a misaligned pair whose $|M\tr|$ assumes the 10th-percentile value resulting from uniformly random node orientations for the given distance \cite{DumphartPIMRC2016}.}
	\label{fig:Misalignment}
	\vspace{-0mm}
\end{figure}

For the two cases of Tx-Rx alignment described above, Fig.~\ref{fig:Misalignment} shows the empirical cumulative distribution function (CDF) of $\eta$ in $\mathrm{dB}$ for different relay densities, which are compared to the respective cases without any relays present. 
With increasing relay density, we observe two distinct effects: firstly, the impact of relative Tx-Rx arrangement decreases and, secondly, the statistical deviation of $\eta$ increases. %because additional relays introduce more randomness to the setup. 
For the coaxial Tx-Rx pair, the observed gains due to relays are either negative or minor, with the median value of $\eta$ even declining with increasing relay density. For the misaligned Tx-Rx case, however, the addition of relays is vastly beneficial: even the low density of $0.1\,\mathrm{relays}/\mathrm{dm}^3$ yields a median gain of $4.7\,\mathrm{dB}$ with $87.1\,\mathrm{\%}$ of network realizations improving $\eta$. These numbers increase to $13.8\,\mathrm{dB}$ and $96.7\,\mathrm{\%}$ for $1\,\mathrm{relay}/\mathrm{dm}^3$. 

We conclude that the presence of randomly arranged passive relays offers the potential to improve the link between a pair of nodes. However, the observed gains are mostly restricted to the mitigation of misalignment losses while considerable gains w.r.t. $\eta$ of a well-aligned Tx-Rx link without relays, as reported for regular setups \cite{shamonina2002magneto,LeeTPE2012}, 
%\textcolor[rgb]{0.92,0.5,0.4}{(Hier wirklich alle verwendeten Referenzen anführen, die das erfuellen)}
do not occur. This is due to the adverse effects described in Sec.~\ref{sec:model}: without any further measures, the probability of getting highly coherent transimpedance phasors in \eqref{eq:SumOfPhasors} is vanishingly low for a randomly arranged relay network (with its $5N$ geometric degrees of freedom from positions and orientations). %In the next section, we overcome the described limitations by enforcing constructive phasors with several parameter adjustments.

The observed effects are analogous to multipath radio channels, where non-coherent addition of paths leads to frequency-selective fading, which is mostly considered adverse in case of a strong line-of-sight (LOS) path. Conversely, multipath is crucial for establishing a link when the LOS path is blocked.

\section{Channel Gain Optimization}
\label{sec:opt}
As discussed in the previous section, the gains brought by randomly arranged relaying networks are limited by the low probability of encountering a large transimpedance. In this section, we propose to overcome this problem in a systematic fashion: we introduce optimization schemes in different parameter domains, namely load switching and frequency tuning, to maximize $\eta$ for a given arrangement with $N$ passive relays. A subsequent performance comparison reveals interesting features of the relaying channel.

\subsection{Load Switching}
The relay loads constitute many degrees of freedom regarding the maximization of $\eta$ when we allow for adaptive loads. Such load adaptation at passive relays has been studied for sum rate maximization in interference radio channels \cite{hassan2015} and for resolving ambiguities in near-field localization \cite{SlottkeASILOMAR2015}. %\cite{Slottke2016}

\noindent Adaptive loads that can achieve any $\tilde{Z}_n \in \mathbb{C}$ would be ideal, but this is not realistic for low-complexity sensor nodes. Instead, we consider open-circuit switching: we assume that every relay is equipped with a switch in series to its resonance capacitor and with logic for opening or closing this switch after receiving a corresponding command. %This requirement could be realized with low hardware and protocol effort. 
As opening the switch of a relay leads to no current flowing in the respective antenna, an open-circuited relay does not interact with other coils.
%As opening the switch blocks current through the relay antenna, an open-circuited relay does not interact with other coils. 

\begin{figure}[!ht]
	\centering
  \psfrag{xaxis}{\raisebox{-.9mm}{\hspace{-2.5mm}\footnotesize{$f$   [$\mathrm{MHz}$]}}}
  \psfrag{yaxis}{\raisebox{1.2mm}{\hspace{-13.5mm}\footnotesize{Channel Power Gain $\eta$ [$\mathrm{dB}$]}}}
 	\includegraphics[width=\linewidth,trim=25 8 42 15,clip=true]{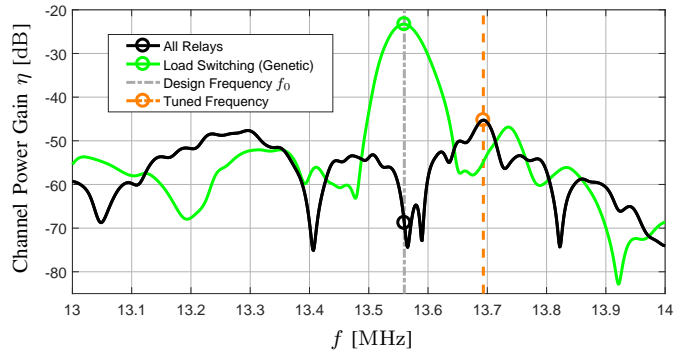}
  \put( -151.5,92.0){\scalebox{.79}{\scriptsize $f_0$}}
 	\caption{Illustration of load switching and frequency tuning in terms of the frequency response of an exemplary passive relaying channel with a density of $10\,$relays/dm$^3$, arbitrary arrangement, and $50\,\mathrm{cm}$ Tx-Rx separation.} 
  \label{fig:Spectrum}
\end{figure}

Each relay being either resonant or open-circuited gives rise to $2^N$ switching states for the relaying network. We want to find the switching state that maximizes $\eta$. Such a binary optimization problem becomes intractable for large $N$, but genetic algorithms are an efficient means to find decent heuristic solutions. %\cite{Davis1991}. 
We therefore employ a genetic algorithm for relay load switching: starting from random switching states, we simulate 500 generations where only the 20 states of largest $\eta$ survive a generation. Per generation, every state produces a mutated child (i.e. with a few switches flipped at random, which may or may not improve $\eta$) while four recombined states are generated to alleviate the problem of local maxima. Fig.~\ref{fig:Spectrum} compares the frequency response of the relaying channel corresponding to an exemplary network for two cases: a setting with all relays in the resonant state, subsequently referred to as the \mbox{all-relays} case, and the same network after genetic load switching optimization which results in a strong peak at $f_0$.

%To study the capabilities of passive relaying networks in resolving misalignment and destructive phasor addition, 
Furthermore, we consider two particularly simple protocols for load switching: A \mbox{$1$-relay} scheme where only the single relay that maximizes $\eta$ is activated (after an exhaustive search over all $N$ relays) and, similarly, an $N\!-\!1$ scheme where 
%only a single relay is deactivated by open-circuiting its antenna to maximize $\eta$, 
only the one relay whose deactivation leads to the greatest $\eta$ is open-circuited, 
while all other relays remain resonant.

\subsection{Frequency Tuning}

We already discussed the frequency selectivity of the passive relaying channel in Section \ref{sec:study}. The proposed scheme chooses the operation frequency at which $\eta$ is maximized while the resonant relay design remains unchanged, as illustrated in Fig.~\ref{fig:Spectrum}. Besides the simplicity of this one-dimensional search, frequency tuning is attractive because of its low hardware complexity: it requires only tunable filters and voltage-controlled oscillators at the Tx and Rx, but no adjustments to the relays.

\subsection{Comparison \& Discussion}

All simulations use the parameters specified in Section \ref{sec:eval}, but now also the Tx and Rx coil orientations are considered uniformly random (equivalent to the relays). The resulting distribution of $\eta$ without relays was investigated in \cite{DumphartPIMRC2016}. We consider a high density of $10\,\mathrm{relays}/\mathrm{dm}^3$ and two different Tx-Rx separations: $15\,\mathrm{cm}$ and $50\,\mathrm{cm}$. 

\begin{figure}[ht!]
	\centering
  \psfrag{EmpCDF}{\raisebox{.8mm}{\hspace{-4.8mm}\footnotesize{Empirical CDF}}}
  \psfrag{CGdB}{\raisebox{-.7mm}{\hspace{-13.7mm}\footnotesize{Channel Power Gain $\eta$ [$\mathrm{dB}$]}}}
	\psfrag{RGdB}{\raisebox{-.7mm}{\hspace{-6.8mm}\footnotesize{Relay Gain [$\mathrm{dB}$]}}}
	\includegraphics[width=\linewidth,trim=34 10 48 24,clip=true]{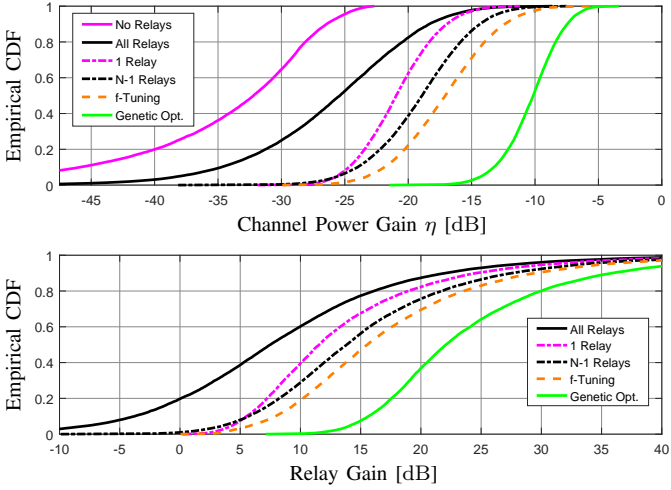}
  \vspace{-5mm}
  \caption{Statistics of $\eta$ for a Tx-Rx pair of $15\,\mathrm{cm}$ separation within a random passive relaying network. We compare unoptimized and optimized relaying networks with a density of $10\,\mathrm{relays}/\mathrm{dm}^3$ to the case without any relays.}
  \label{fig:OptPerformance15cm}
\end{figure}

\begin{figure}[ht!]
  \psfrag{EmpCDF}{\raisebox{.8mm}{\hspace{-4.8mm}\footnotesize{Empirical CDF}}}
  \psfrag{CGdB}{\raisebox{-.7mm}{\hspace{-13.7mm}\footnotesize{Channel Power Gain $\eta$ [$\mathrm{dB}$]}}}
	\psfrag{RGdB}{\raisebox{-.7mm}{\hspace{-6.8mm}\footnotesize{Relay Gain [$\mathrm{dB}$]}}}
	\includegraphics[width=\linewidth,trim=34 10 48 24,clip=true]{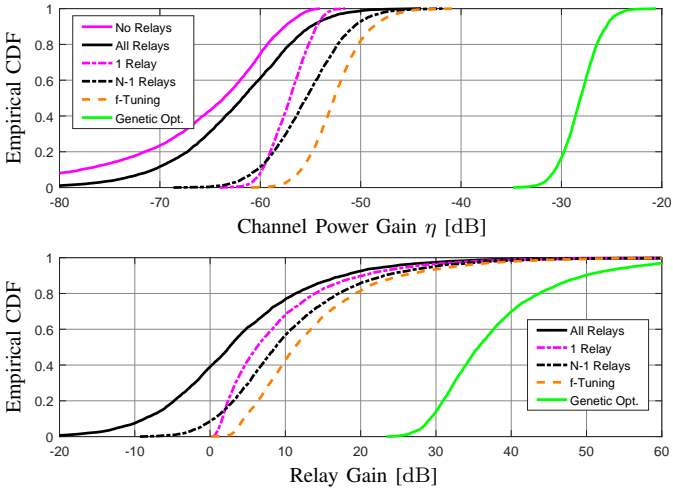}
  \vspace{-5mm}
  \caption{Statistics of $\eta$ for a Tx-Rx pair of $50\,\mathrm{cm}$ separation within a random passive relaying network.}
	\label{fig:OptPerformance50cm}
\end{figure}

The results for $15\,\mathrm{cm}$ and $50\,\mathrm{cm}$ Tx-Rx separation are shown in Fig.~\ref{fig:OptPerformance15cm} and \ref{fig:OptPerformance50cm}, respectively, where we compare the no-relays and all-relays cases to the proposed optimization schemes. Empirical CDFs are shown for $\eta$ and the relay gain, which is the ratio of $\eta$ of a specific scheme to $\eta$ of the \mbox{no-relays} link with the same individual Tx-Rx arrangement.

The performance of the all-relays case agrees with high-density results of Sec.~\ref{sec:eval}. Strictly positive relay gains are realized by frequency tuning, genetic load switching, and the $1$-relay scheme. The $\eta$-statistics of the $1$-relay scheme show that robust mitigation of misalignment losses is possible with a single relay. 
In contrast, the key feature of the $N-1$ scheme is the ability to mitigate destructive superposition in the transimpedance term \eqref{eq:SumOfPhasors} by deactivating a detrimental relay. 
Therewith, the $N-1$ scheme outperforms the all-relays case by a considerable margin, but its single degree of freedom is insufficient to achieve highly coherent phasors. Frequency tuning robustly finds reasonably constructive conditions in the frequency domain and outperforms the $1$-relay and $N\!-\!1$ cases. However, we observe that frequency tuning has limited optimization potential as well and, in consequence, fails to achieve the gains that are possible at high relay densities. Load switching with the genetic algorithm, however, shows great performance and outperforms the other discussed schemes by orders of magnitude. The scenario of $15\,\mathrm{cm}$ separation in Fig.~\ref{fig:OptPerformance15cm} exhibits a median $\eta$ of $-32.1\,\mathrm{dB}$ without relays, which is improved to $-10.1\,\mathrm{dB}$ by relays and genetic load switching. This would enable power transfer with an efficiency of about $10\,$\%. In the $50\,\mathrm{cm}$ scenario of Fig.~\ref{fig:OptPerformance50cm}, minimum and median relaying gains of $23.4\,\mathrm{dB}$ and $35.6\,\mathrm{dB}$, respectively, are achieved with genetic load switching.
%The CDFs of the proposed optimization schemes show a behavior similar to channel hardening \textcolor[rgb]{0.92,0.5,0.4}{(Refer to respective figure, this can not be observed for all "CDFs of proposed optimization schemes")}: the diversity provided by the high relay density allows the schemes to robustly achieve a certain link quality, thus $\eta$ exhibits little deviation.

Finally, we compare the all-relays case to genetic load switching in terms of narrowband communication rates for the described $50\,\mathrm{cm}$ setup with $P\t = 1\,\mu\mathrm{W}$ and a bandwidth of $B = 1\,\mathrm{kHz}$. We consider thermal noise in the Rx circuit as the only limitation to rates (analysis of the system model showed that induced Rx currents due to fields generated by thermal noise at other nodes are negligible, even for high-density networks). The achievable rate is calculated as $B \log_2(1+\eta P\t / P_\mathrm{N})$ with noise power $P_\mathrm{N} = k_\mathrm{B} T B \cdot \mathrm{NF}$ where $k_\mathrm{B}$ is the Boltzmann constant. We assume a temperature of $T = 300\,\mathrm{K}$ and a noise figure of NF $= 15\,\mathrm{dB}$ at the Rx. 

\begin{figure}[!ht]
	\centering
  \psfrag{RDensUnit}{\raisebox{-0.9mm}{\hspace{-12.0mm}\footnotesize{Relay Density [$\mathrm{relays}/\mathrm{dm}^3$]}}}
  \psfrag{Rbps}{\raisebox{0.6mm}{\hspace{-5.5mm}\footnotesize{Rate [$\mathrm{bps}$]}}}
  \includegraphics[width=\linewidth,trim=30 0 45 18,clip=true]{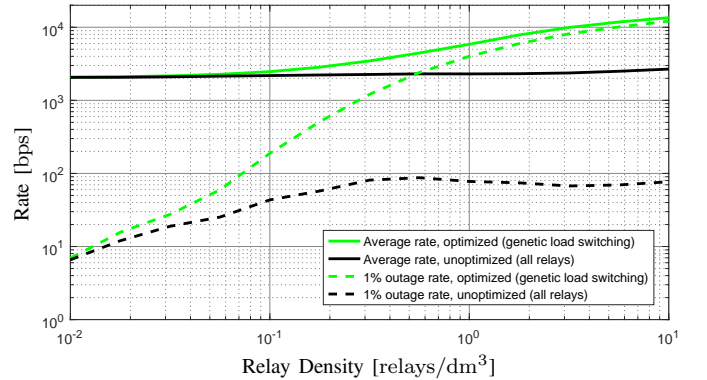}
  \caption{Narrowband rates for $50\,\mathrm{cm}$ Tx-Rx separation and randomly arranged relaying networks, comparing all-relays case and genetic scheme. We assume $P\t = 1\,\mu\mathrm{W}$ and $B = 1\,\mathrm{kHz}$ with thermal noise at $300\,\mathrm{K}$ and NF $= 15\,\mathrm{dB}$.}
  \label{fig:Rates}
\end{figure}

Fig.~\ref{fig:Rates} shows the evolution of average rates and 1\% outage rates over relay density for the all-relays case and the genetic scheme. In addition to gains in terms of average rate, we observe considerable robustness improvements through optimization due to channel hardening: the outage rate approaches the average rate because the diversity offered by a high-density relaying network enables the genetic load switching scheme to robustly achieve high rates. In the all-relays case, we observe an increasing outage rate for low relay densities, which shows that the misalignment mitigation aspect outweighs destructive effects of passive relays in terms of robustness (cf. Sec.~\ref{sec:eval} and Fig.~\ref{fig:Misalignment}, where the outcome of this trade-off was unclear). 
The rates becoming almost constant for higher densities in the all-relays case is consistent with the large-density limiting distribution of $\eta$ observable in Fig.~\ref{fig:Misalignment}.

\section{Summary and Conclusions}
In the context of inductively coupled networks, we extend the well-known approach of MI passive relaying to networks of arbitrary geometries. In networks with random relay arrangements, such as can be expected for ad-hoc sensor networks, we show that the relays can mitigate misalignment losses between a Tx-Rx pair, but do not reach their full potential of performance gains observed in literature for carefully chosen relay placements. To better utilize a set of randomly arranged relays we propose and evaluate two low-complexity optimization schemes based on binary load switching and frequency tuning. Both methods show significant improvements in power gain with a median improvement of up to $35.6\,\mathrm{dB}$ compared to non-optimized settings. The proposed methods are therefore a promising means to enable reliable communication or efficient power transfer in dense inductively coupled ad-hoc networks.

\appendices
%\numberwithin{equation}{section}%
\section*{Appendix\\ Power Gain through passive, reciprocal Two-port}\label{app:match}%
We derive a general formula in $Z$-parameters for power gain $\eta$ over passive, reciprocal two-port networks, e.g. for $\Z$ in \eqref{eq:EquivTwoport}. This is achieved with simultaneous conjugate matching, i.e. when source and load impedance are the complex conjugates of the respective port impedances like in Fig.~\ref{fig:AppMatch}.
\newcommand{\AppFigVSpace}{\vspace{-4mm}}
\vspace{-4mm}

\begin{figure}[ht]
   \centering
    \subfloat[Simultaneous conjugate matching of a passive, reciprocal two-port network.]{
      \psfrag{vT}{\hspace{1.5mm}\raisebox{.7mm}{$v\t$}}
      \psfrag{p}{\footnotesize\hspace{.5mm}\raisebox{.3mm}{$+$}}
      \psfrag{m}{\footnotesize\hspace{.5mm}\raisebox{.3mm}{$-$}}
      %\psfrag{ZiC}{\hspace{-4.5mm}\raisebox{1mm}{$Z\t \!\overset{!}{=}\! Z_\text{in}^*$}}
      \psfrag{ZiC}{\hspace{2.85mm}\raisebox{1mm}{$Z_\text{in}^*$}}
      \psfrag{Zi}{\hspace{-.1mm}\color{red}\raisebox{-.1mm}{$Z_\text{in}$}}
      \psfrag{mtx}{\hspace{-3.63mm}\small\raisebox{1.5mm}{$\mtx{ll}{\!\!\!Z_{1,1}\!\!\! & \!\!Z_{2,1}\!\!\!\! \\ \!\!\!Z_{2,1}\!\!\! & \!\!Z_{2,2}\!\!\!\!}$}}
      \psfrag{Zo}{\hspace{-.5mm}\color{red}\raisebox{-.1mm}{$Z_\text{out}$}}
      \psfrag{ZoC}{\hspace{-1.3mm}$Z_\text{out}^*$}
      \hspace{1mm}
      \includegraphics[height=2.03cm,trim=30 0 35 2,clip=true]{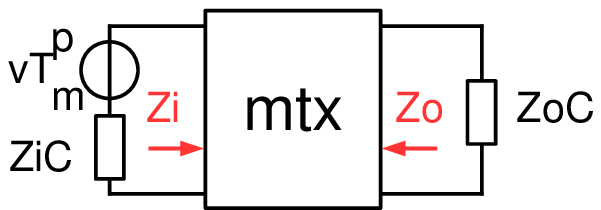}
      \label{fig:AppMatch}
    }\qquad
    \subfloat[Load-side Thevenin equivalent circuit.]{
      \centering
      \psfrag{vT}{\hspace{1.2mm}\raisebox{.7mm}{$v\t'$}}
      \psfrag{p}{\footnotesize\hspace{.5mm}\raisebox{.3mm}{$+$}}
      \psfrag{m}{\footnotesize\hspace{.5mm}\raisebox{.3mm}{$-$}}
      \psfrag{Zo}{\hspace{1.2mm}\raisebox{1mm}{$Z_\text{out}$}}
      \psfrag{ZoC}{\hspace{-1.3mm}$Z_\text{out}^*$}
      \hspace{5mm}
      \includegraphics[height=2.1cm,trim=30 1 35 0,clip=true]{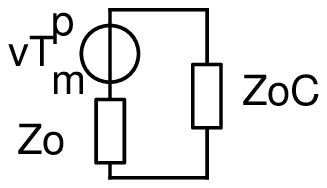}
      \hspace{2mm}
      \AppFigVSpace
      \label{fig:AppThevenin}
    }
   \caption{General circuit and load-side equivalent circuit.}
\end{figure}

%\begin{figure}[!ht]
    %\centering
    %\subfloat[Ansa.\label{fig:Ansa}]{%
      %\includegraphics[width=.45\columnwidth,trim=0 0mm 0 3mm,clip=true]{./Appendix-Simultaneous-Conj-Match}
    %}
    %\hfill\vspace{-3mm}
    %\subfloat[Zwara.\label{fig:Zwara}]{%
      %\includegraphics[width=.45\columnwidth,trim=0 0mm 0 .5mm,clip=true]{./Appendix-Thevenin}
    %}
    %\caption{Keepo.}
    %\label{fig:Hey}
%\end{figure}

\balance
Writing the port impedances of the loaded two-port network and enforcing conjugate matching conditions corresponds to
\vspace{2.4mm}
\begin{align}
& Z_\text{in} \overset{!}{=} Z_{1,1} \!-\! \f{Z_{2,1}^2}{Z_{2,2} \!+\! Z_\text{out}^*} \ , &
& Z_\text{out} \overset{!}{=} Z_{2,2} \!-\! \f{Z_{2,1}^2}{Z_{1,1} \!+\! Z_\text{in}^*} \ ,
\end{align}
%i.e. two complex equations in the unknown port impedances $Z_\text{in}, Z_\text{out}$, or rather four equations in their real and imaginary parts. Lengthy application of basic algebra gives the solution
i.e. two complex equations in $Z_\text{in}, Z_\text{out}$. A solution thereof is
%(Z_{1,1} - Z_\text{in})(Z_{2,2} + Z_\text{out}^*) &= Z_{2,1}^2 \\
%(Z_{1,1} + Z_\text{in}^*)(Z_{2,2} - Z_\text{out}) &= Z_{2,1}^2
%\end{align}
%with solution %(which can be obtained by considering four equations in real and imaginary parts)
\begin{align}
Z_\text{in}  &= \Big( \!\sqrt{1\!-\!\rho^2} \sqrt{1\!+\!\chi^2} - j\rho\chi \Big) \re Z_{1,1} + j\im Z_{1,1} \\
Z_\text{out} &= \Big( \!\sqrt{1\!-\!\rho^2} \sqrt{1\!+\!\chi^2} - j\rho\chi \Big) \re Z_{2,2} + j\im Z_{2,2}
\end{align}
where $\rho+j\chi = Z_{2,1} \big/ \!\sqrt{\re Z_{1,1} \re Z_{2,2}}$, cf. \eqref{eq:alphabeta}. Other solutions are either meaningless or unphysical due to zero or negative real parts. 
Now $v\t' = v\t \cdot Z_{2,1} \, / \, (Z_{1,1} + Z_\text{in}^*)$ of the Rx-side equivalent circuit in Fig.~\ref{fig:AppThevenin} can be rearranged to
\begin{align}
\f{v\t' \, \big/ \sqrt{4\re Z_\text{out}}}{v\t \, \big/ \sqrt{4\re Z_\text{in}} \ } = \f{\rho + j\chi}{1 + \sqrt{1\!-\!\rho^2} \sqrt{1\!+\!\chi^2} + j\rho\chi} =: h \ .
\label{eq:apph}
\end{align}
Note the phase information in the channel coefficient $h \in \mathbb{C}$. Its squared absolute value is the ratio of active powers through Rx and Tx ports $\eta = \f{P\r}{P\t} = |h|^2$, which leads to the power gain formula \eqref{eq:PowerGain} after basic algebraic manipulations. \hfill$\square$

%Result \eqref{eq:apph} can also be obtained by lengthy rearrangement of generalized $S$-parameter results \cite{Rahola2008}. In fact, $h$ is identified as $S_{21}$ of scattering matrix $\textbf{S} = (\textbf{A} - \gamma\eye_2)(\textbf{A} + \gamma\eye_2)^{-1}$ with
%\begin{align}
%\textbf{A} &= \mtx{cc}{\!\!1+j\rho\chi\!\! & \!\!\rho+j\chi\!\! \\ \!\!\rho+j\chi\!\! & \!\!1+j\rho\chi\!\! } , &
%\!\! \gamma &= \sqrt{1\!-\!\rho^2}\sqrt{1\!+\!\chi^2} \ .
%\end{align}
The left-hand side of \eqref{eq:apph} being a ratio of power waves suggests the use of generalized $S$-parameters. In fact, after rearrangement of $S$-parameter results \cite{Rahola2008}, we identify $h$ as $S_{21}$ of scattering matrix $\textbf{S} = (\textbf{A} - \gamma^*\eye_2)(\textbf{A} + \gamma\eye_2)^{-1}$ with
\begin{align}
\textbf{A} \!&=\! \mtx{cc}{1 & \!\!\!\rho+j\chi\!\!\! \\ \!\!\rho+j\chi\!\!\! & 1} , &
\!\! \gamma \!&=\! \sqrt{1\!-\!\rho^2}\sqrt{1\!+\!\chi^2} + j\rho\chi \ .
\end{align}

\balance

\IEEEtriggeratref{0}
\bibliographystyle{IEEEtran}
\bibliography{IEEEabrv,BibGregor}

\end{document}